((please add manuscript number))

# Uniform 100mm wafer-scale synthesis of graphene on evaporated Cu (111) film with quality comparable to exfoliated monolayer

Li Tao,[1] Milo Holt,[1] Jongho Lee,[1] Harry Chou,[2] Stephen J. McDonnell,[3] Domingo A. Ferrer,[1] Matías Babenco,[4] Robert M. Wallace,[3] Sanjay K. Banerjee,[1] Rodney S. Ruoff,[2] Deji Akinwande [1,*]

[1]Microelectronics Research Center, Department of Electrical Engineering,

[2]Department of Mechanical Engineering and the Texas Materials Institute,

The University of Texas at Austin, Austin, TX 78758

[3]Department of Materials Science and Engineering,

The University of Texas at Dallas, Richardson, TX 75080

[4]Department of Electrical Engineering, Universidad Nacional de Córdoba, Córdoba, Argentína

*Abstract:* Monolayer graphene has been grown on crystallized Cu (111) films on standard oxidized Si 100 mm wafers. The monolayer graphene demonstrates high uniformity (>97% coverage), with immeasurable defects (>95% defect-negligible) across the entire wafer. Key to these results is the phase transition of evaporated copper films from amorphous to crystalline at the growth temperature as corroborated by X-ray diffraction and electron backscatter diffraction. Noticeably, phase transition of copper film is observed on technologically ubiquitous oxidized Si wafer where the oxide is a standard amorphous thermal oxide. Ion mass spectroscopy indicates that the copper films can be purposely hydrogen-enriched during a hydrogen anneal which subsequently affords graphene growth with a sole carbonaceous precursor for low defect densities. Owing to the strong hexagonal lattice match, the graphene domains align to the Cu (111) domains, suggesting a pathway for increasing the graphene grains by maximizing the copper grain sizes. Fabricated graphene transistors on a flexible polyimide film yield a peak carrier mobility ~4,930 cm$^2$/Vs.

*Communication*

The wafer-scale synthesis of monolayer graphene with complete surface coverage and low defects on Si wafers for direct integration with standard CMOS processes is a necessary requirement for very large-scale graphene nanoelectronics.[1-6] To this end, the Cu (111) surface has been identified by some as the preferred catalytic metal with a good lattice matching (lattice mismatch<4% at 300ºC) to graphene is essential for achieving low defects with high uniformity.[7-9] However, Cu (111) films are typically obtained on single crystal epitaxial substrates such as sapphire,[7, 10] which do not offer the low-cost and industrial scale processing of standard Si substrates. In contrast, copper films deposited on thermal oxides on Si (thus, $SiO_2$/Si) are mostly amorphous as-deposited, and usually form polycrystalline grains with random orientations after thermal treatment. Although the Cu (111) facet is energetically favorable considering its minimum surface energy compared to other facets,[11-12] it is not straightforward to obtain the (111) texture surface due to competing factors such as strain energy[12-13] and restricted boundaries.[14-15] The influence of these factors result in an increased fraction of (200),[14] or (220),[15] facets that are said to lead to non-uniform quality and substantial defects in the synthesized graphene.[8]

In this work, we report both the phase transition of evaporated amorphous copper to crystallized Cu (111) films on 100-mm $SiO_2$/Si wafers after very high flow-rate $H_2$ thermal anneal at 900-1000ºC, and growth of graphene by chemical vapor deposition (CVD) on the annealed substrate. Electron back-scatter diffraction (EBSD) reveals that the crystallized copper film affords a (111) orientation in ~97% of the characterized surface. Time-of-flight secondary ion mass spectroscopy (TOF-SIMS) indicates hydrogen enrichment of the copper film during the high $H_2$ anneal, and we have previously reported this contributes to the quality of the monolayer graphene.[16] Raman mapping of the CVD graphene shows high uniformity across the wafer with an average intensity ratio of the 2D-peak to G-peak ($I_{2D}/I_G$) ~3.2, average full width at half maximum of the 2D-peak ($FWHM_{2D}$) ~30 cm$^{-1}$, and with very low defect density as measured by the D-peak to G-peak intensity ratio ($I_D/I_G$) which is <0.2 for over 95% of the surface; all these observations indicate that high quality graphene comparable to exfoliated graphene has been obtained.[17-18] These results pave the path for wafer-scale graphene nanotechnology with near-perfect monolayer coverage and with the material quality to afford the high-yield essential for very large-scale integrated (VLSI) systems.

Typical synthesis process in this study, as depicted in **Figure** 1, included satuated hydrogen annealing step, growth step with ultra-high purity methane only and a two-step cooling before unloading samples from a vertical cold-wall chamber with separate showerhead and substrate heaters (see *Experimental* section, *Graphene synthesis*). The saturated hydrogen annealing step was found to be critical for achieving the results reported in this work, as discussed further below. Immediately after the synthesis of graphene, the sample was characterized with Raman spectrocope using a blue laser as light source (see *Experimetnal* section, *Material characterization*). Raman spectra from 5 spot-locations over a 100-mm diameter wafer as denoted in **Figure 2**a (**B** for bottom, **C** for center, **L** for left, **R** for right, and **T** for top, respectively) are presented in Figure 2b. The average FWHM$_{2D}$ in Figure 2b is ~28 cm$^{-1}$, with $I_{2D}/I_G$ ~3, and no measurable D-peak signifying the successful growth of high-quality monolayer graphene. To verify the uniformity of the synthesized graphene, Raman maps across 200 × 200 μm$^2$ areas were obtained with each centered on the locations (**B**, **R**, **L**, and **T**) mentioned above, and evaluated as $I_{2D}/I_G$ and $I_D/I_G$ ratio maps in Figures 2c and 2d, respectively. For instance, the histogram data in Figures 2e and 2f indicate average values of 3.2±0.19 for $I_{2D}/I_G$ and that >95% of the scanned area has $I_D/I_G$ <0.2. Based on the Raman characterization, monolayer graphene has been achieved on a 100-mm wafer scale with negligible or no measurable defects.

The successful synthesis of uniform monolayer graphene with negligible disorder depends on using a hydrogen-enriched evaporated Cu (111) film as demonstrated in our previous study.[16] The high percentage (over 96.8% in a 1×1 mm$^2$ area) for a Cu (111) orientation was observed by EBSD mapping (**Figure 3**a) after annealing and then graphene synthesis. This distinguishes the evaporated copper film reported here from conventional copper foils, which show a Cu (100) orientation as the dominant facet after the same annealing process (see supporting information **Figure S1**). X-ray diffraction (XRD) shown in Figure 3b also shows a dominant (111) orientation. Cu (111) surface offers some advantages for growing high quality graphene relative to other facets as also observed in other reports.[7,8,14,15] A microscopic route for graphene growth on copper catalytic surface includes: i) adsorption of precursor such as $CH_4$ and its decomposition into carbon monomer/dimer,[19-21] ii) the diffusion of carbon monomers/dimers leading to the formation of clusters,[20-21] and iii) the attachment (growth) or detachment (etching) of carbon at the edge of existing cluster/nuclei. First, the adsorption energy for the initial steps of decomposing $CH_4$ on Cu (111) is lower than other facets.[19-20] Secondly, the diffusion rate of a carbon monomer/dimer on Cu (111) is higher than other facets like Cu (100).[8, 20] In addition, the Cu (111) surface offers less nuclei density and a faster growth rate[21] for graphene grains than other crystal facets, which altogether could yield large graphene domains under the conditions reported here and even at temperatures $\leq 900$ °C.[16]

The negligible or weak D-peak intensity in the Raman spectroscopy of graphene grown on Cu (111) film is likely due to grain boundaries that are dominated by zig-zag (ZZ) edges. It has been previously reported that the root causes of defects monitored by the D-peak in the Raman spectroscopy of otherwise clean graphene can be attributed to two kinds of imperfections that break translational symmetry: i) graphene domains or grain boundaries with mixed zig-zag and arm-chair (AC) edges,[22-24] and/or ii) the boundary between graphene and imperfect catalytic surface or underlying substrate.[25-27] Recent density functional theory (DFT) modeling of the Cu (111) surface suggested that the dominant growing edge of graphene should be in the ZZ direction.[28-30] The absence of AC edges was said to be because of their rapid passivation by copper atoms.[30] Graphene domains dominated by ZZ edges/boundaries show no or negligible D-peak in the Raman spectra, whereas domains with combined AC and ZZ edges will induce a significant D-peak intensity.[23] For this reason, the polycrystalline graphene on evaporated copper film observed in this work should have grain boundaries primarily composed of ZZ edges based on the negligible D-peak observed in the Raman spectroscopy.

Another important synthesis feature is the role of $H_2$ on the annealed Cu (111) film which contrasts with conventional foils. Owing to the higher adsorption of hydrogen on Cu (111) than other crystal orientations,[31-33] a significant amount of hydrogen can adsorb and diffuse into the Cu (111) film during hydrogen saturation annealing (see supporting information **Figure S2**). The absorbed hydrogen can subsequently diffuse to the film surface to serve as a co-catalyst for monolayer graphene growth,[34] thereby eliminating the need for a gas-phase $H_2$ precursor during the growth step, which can result in graphene etching with a noticeable D-peak.[16, 35] On the other hand, the presence of hydrogen during the growth process likely contributes to an increase in the average copper grain size. For instance, the average grain size of ~15-20 μm obtained without $H_2$ can further be increased to 20-25 μm with $H_2$ precursor during growth as shown in Figures 3c and 3d.

Samples with a graphene monolayer synthesized on copper film using the route reported here, albeit on a limited size scale (~5×7.6 cm$^2$, **Figure 4**a), have been transferred to various substrates such as $SiO_2$/Si, quartz, and flexible polyimide sheets via a two-step etching procedure (see *Experimental* section, *Graphene transfer*). It is worthwhile to note that the etching time for evaporated copper film is greatly reduced (~5 mins) compared to conventional copper foils that are more than an order of magnitude thicker. In our transfer process, the PMMA was dissolved away in a 50 °C acetone bath for ~30 mins without a post-transfer baking step that is commonly reported in the literature but was also observed to cause residues of PMMA on the graphene surface (Figure 4b).[36-37] Figure 4c shows an optical image of synthesized graphene that has been transferred onto a standard 300-nm thick $SiO_2$/Si with our two-step transfer process. A typical thickness of the graphene layer is ~0.6 nm measured by atomic force microscopy (AFM). The histogram of the FWHM of the Raman 2D peak (Figure 4d) indicates the transferred film is monolayer graphene. A scanning tunneling microscope (STM) image of graphene on the copper film prior to the transfer (**Figure 5**a), together with a high resolution cross-sectional transmission electron microscope (TEM) image of the transferred graphene (Figure 5b), confirm that the graphene film was monolayer throughout the whole process. We suggest that the results here are of higher quality in terms of $I_D/I_G$ and $FWHM_{2D}$, and also offer cost and scale advantages, compared to those achieved with epitaxial copper on expensive non-Si substrates.[7, 9-10] Raman signatures of the graphene in this work are compared in **Table 1** to those for graphene obtained by other methods [7, 9-10, 17-18, 38-47].

High quality graphene monolayer on copper film was also transferred to polyimide plastic sheets,[48] using the two-step etching wet-transfer, for the fabrication of flexible graphene field-effect transistors (GFET). The device fabrication followed the procedure in *Experimental* section *Device fabrication*. Electrical characterization of the embedded-gate GFETs exhibited the typical V-shape profile in the $I_D$-$V_G$ (Figure 5c). Increased mobility for electron was observed with shorter channel (Figure 5d), indicative of less scattering of the charge carriers. A peak electron mobility of ~4,900 cm$^2$/Vs had been obtained as in our previous report,[48] which is 5-10 times larger than previously reported values for GFETs fabricated directly on flexible substrates.[49-50]

In summary, chemical vapor deposition of high quality monolayer graphene on copper film evaporated onto 100-mm SiO$_2$/Si wafers has been demonstrated. TEM, STM, and Raman mapping across the wafer showed that uniform monolayer graphene with negligible defects were successfully synthesized. The high quality uniform wafer-scale monolayer synthesized in this study results from the unique growth on hydrogen-rich annealed Cu (111) films as corroborated by EBSD, XRD, and TOF-SIMS. A time-efficient two-step solution-based transfer of the graphene film was realized that preserves the high quality monolayer character of the graphene after the transfer process. Compared to conventional copper foils, this approach has the potential for direct integration with standard CMOS processes either using a transfer-free process[51] or by direct bonding to a target Si wafer which is currently a matter of further research. The quality of the synthesized graphene on copper film achieved in this work exceeds that from epitaxial copper substrates and is comparable to exfoliated graphene but with scalability beyond the reach of exfoliation methods.

*Experimental*

*Graphene synthesis:* As depicted in Figure 1, the synthesis procedure began from the electron-beam evaporation of 0.5-1 µm copper (Plasmaterial®, 99.99% pellets) film at $10^{-6}$ Torr on a 300nm $SiO_2$/Si wafer. Evaporated copper film sample was then annealed in a vertical cold-wall chamber (Aixtron® Black Magic Nanoinstruments) with separate showerhead and substrate heaters under a hydrogen-saturated environment at a typical temperature of 975 °C for 5 min. Immediately after the annealing step, hydrogen was purged from the chamber and ultra-high purity methane (99.999% Matheson®) with typical flow rates of 5-10 sccm was circulated for 5 min for graphene synthesis. After growth, the chamber was cooled from the growth temperature to 550 °C at a rate of 50 °C/min in a gas-free chamber. The heaters were then turned off and cooling continued with 500 sccm of $N_2$ gas. The samples were retrieved from the chamber at temperatures below 120 °C.

*Material characterization:* Renishaw® In-Via Raman Microscope with He-Cd blue laser (442 nm wavelength) was employed to directly monitor the quality of graphene grown on the copper film [16]. Raman mapping data was analyzed and plotted using MATLAB® program. Electron back scattering diffraction (EBSD) was performed on EDAX/TSL OIM® collection system attached to Zeiss® Neon 40 scanning electron microscope and analyzed with MATLAB® for digitized inversed pole figure. X-ray diffraction was performed on a Philips® X'Pert Pro X-ray system, and depth profiling of hydrogen was carried out on an IONTOF® GmbH time-of-flight secondary ion mass spectroscope with 1 nm depth resolution. Veeco® tapping-mode atomic force microscope were used for morphology and surface analysis. TECNAI G2 F20 X-TWIN transimission electron microscope was used for cross-sectional image of transferred graphene sandwiched between $SiO_2$/Si and epoxy.

*Graphene transfer:* The graphene on copper film was spun coated with 200 nm PMMA first and dried in vacuum (30 mbar) for 8 hrs. Copper film provides smoother surface than foil, thus decreasing roughness induced wrinkles during transfer/multi-layer stacking of graphene film. The PMMA/graphene/Cu was then released from the $SiO_2$/Si substrate after wet etching of $SiO_2$ in buffered oxide etchant (BOE, 6:1), and subsequently placed in an ammonium persulfate (APS) solution to etch the copper followed by DIW cleaning. The cleaning was repeated for five more cycles. The floating PMMA/graphene film was then transferred onto the target substrate and vacuum dried (30 mbar). Finally, the PMMA was dissolved away in a 50 °C acetone bath for ~30 mins without a post-transfer baking step.

*Device fabrication:* The embedded gate was pre-patterned on polyimide sheets by e-beam lithography (EBL) on PMMA, and metal lift-off of (5 nm Ti plus 45 nm Pd or Au), followed by atomic deposition of 15 nm $Al_2O_3$ as gate dielectrics. After the transfer of graphene onto pre-patterned polyimide substrate, the channel region was defined with EBL followed by oxygen plasma etching at 50 W, 200 mTorr for 50 s. At last, source and drain contacts were formed by EBL and metail lift-off with 0.5 nm Ti plus 50 nm Pd to complete the device fabrication. All EBL work here were performed on a JEOL® 6000 e-beam machine at 50 kV on 200 nm thick PMMA that is coated with contuctive polymer E-SPACER® 300Z.

*Acknowledgements*
The authors thank A. Lee, Dr. Y. Hao, and Dr. H. Li for insightful discussions. Graphene synthesis was performed in an Aixtron® Black Magic system. This work is supported in part by the Nanoelectronic Research Initiative (NRI SWAN Center), and the Office of Naval Research under the program of Dr. Chagaan Baatar.

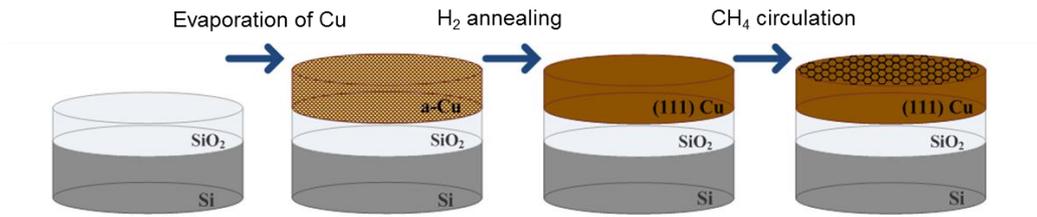

**Figure 1.** Process flow of chemical vapor deposition of graphene on evaporated Cu (111) film

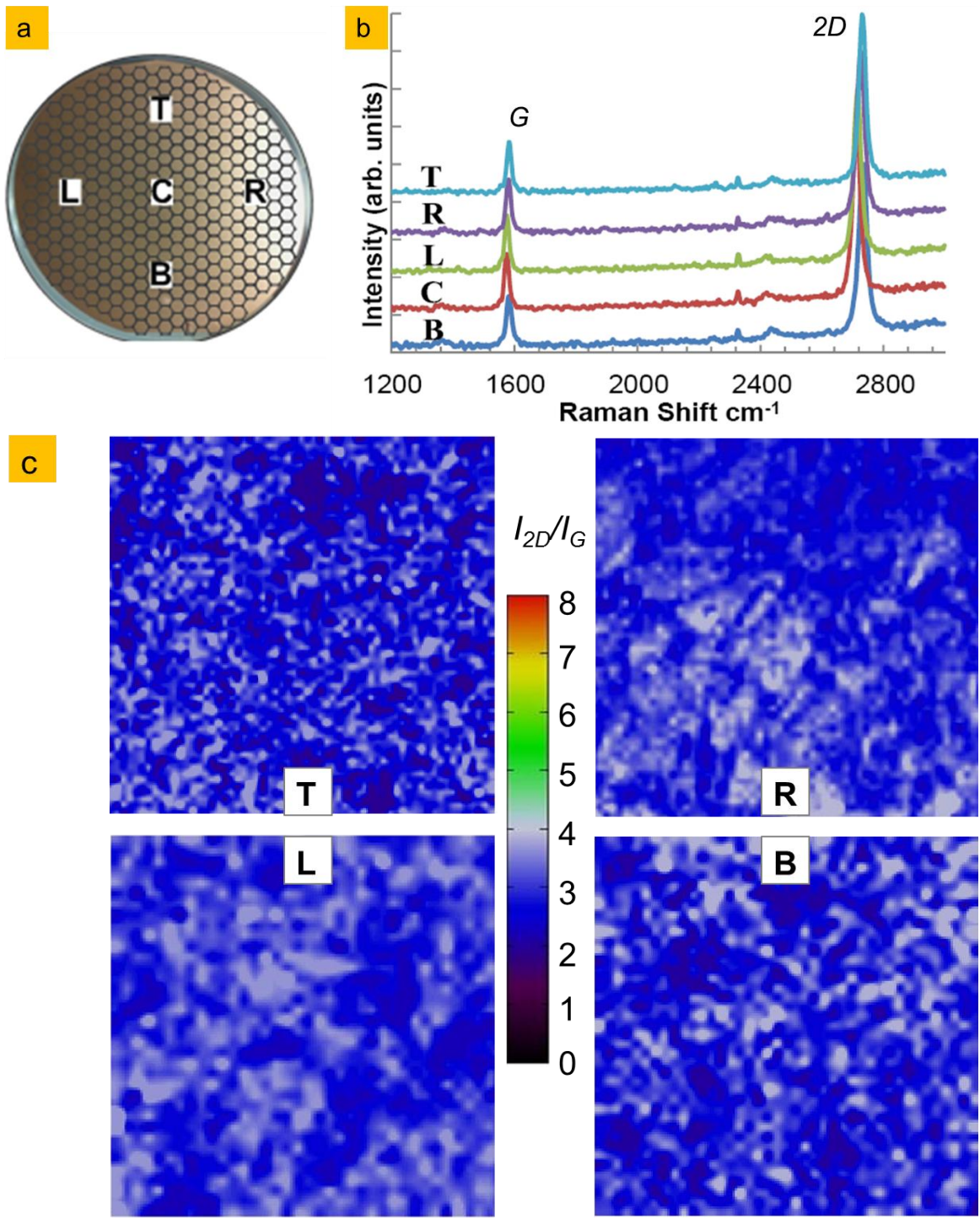

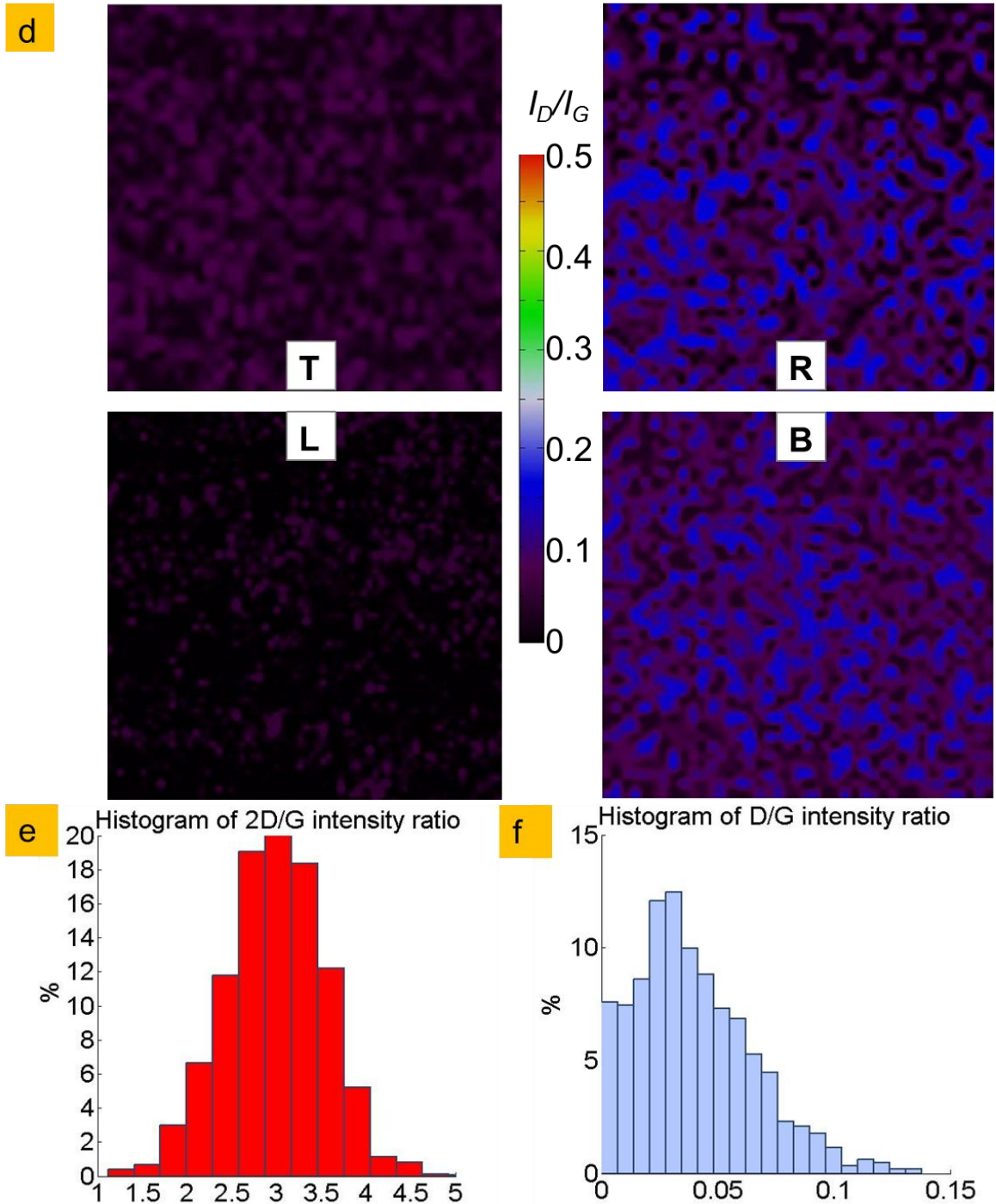

**Figure 2.** Raman characteristics of synthesized graphene: a) Illustration of graphene on an optical image of Cu/SiO$_2$/Si 100-mm wafer; b) spot scans of 5 locations indicated on the wafer image: bottom (**B**), center (**C**), left (**L**), right (**R**) and top (**T**); c,d) ratio maps of $I_{2D}/I_G$ and $I_D/I_G$ from Raman scan on 4 locations (scan size 200×200 μm$^2$); e,f) histograms showing monolayer graphene with an average $I_{2D}/I_G$ ~3, and negligible D-peak (indicative of defects or disorder) with $I_D/I_G$ <0.2 for over 95% of the scanned area.

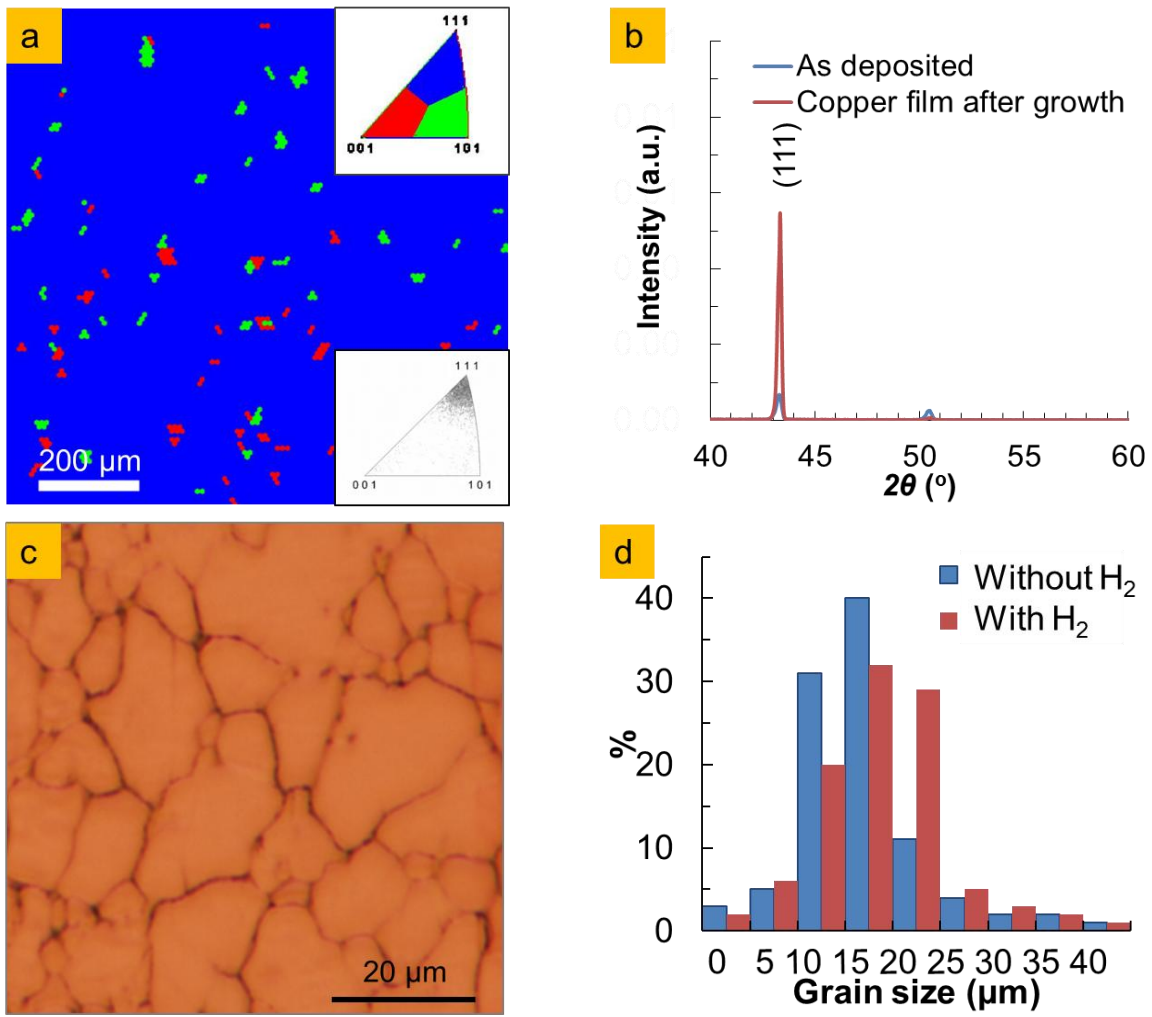

**Figure 3.** a) Electron back-scatter diffraction mapping revealing the dominant Cu (111) crystal orientation after graphene growth. The insets are the inverse pole legend (top) and raw data (bottom); b) X-ray diffraction of copper film before and after the synthesis of graphene. The dominant orientation observed after growth is Cu (111) at $2\theta \sim 43.3^{\circ}$; c) optical image showing copper grains with an average size ~18 μm; d) histograms of copper grain size with and without $H_2$ flow during graphene growth.

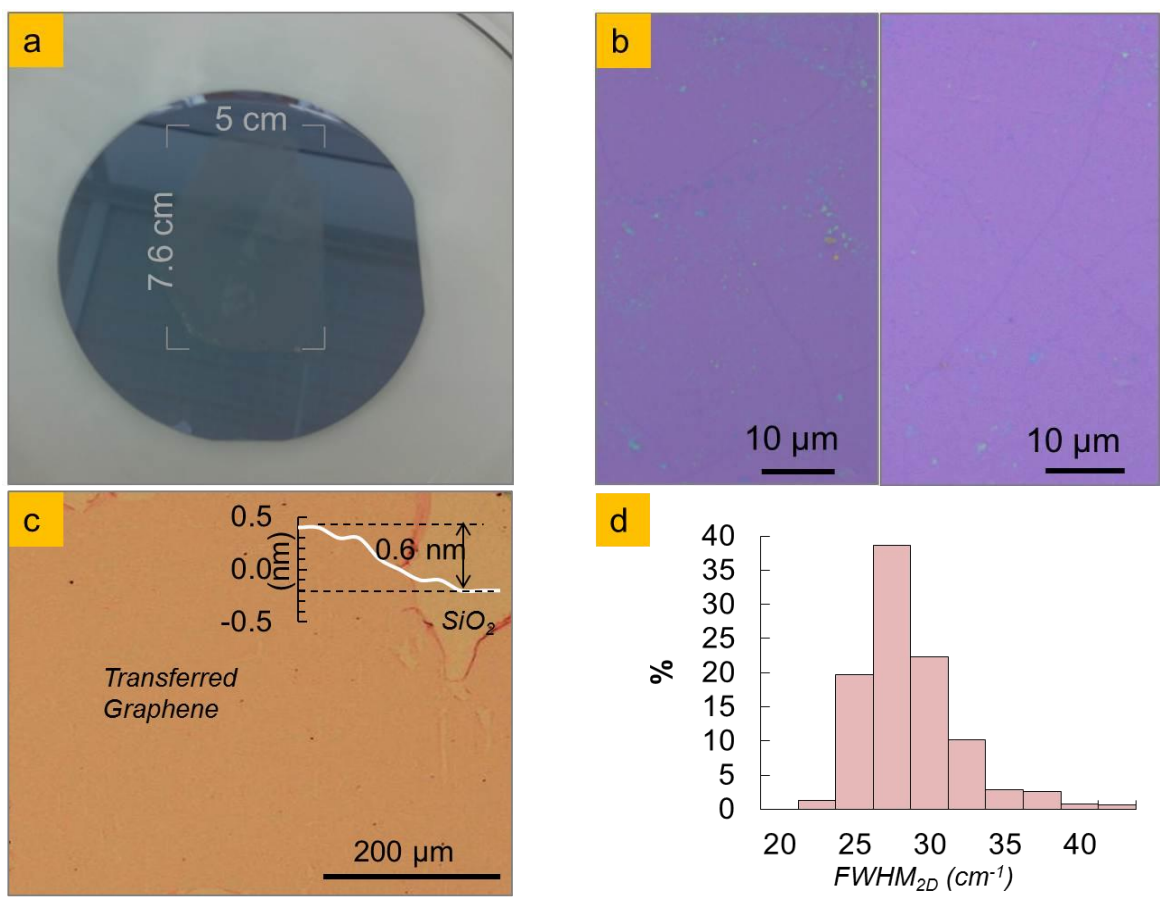

**Figure 4.** a) PMMA/Graphene film (~5×7.6 cm$^2$) floating above a target 100-mm wafer during transfer process; b) Graphene film transferred onto SiO$_2$/Si substrate: the left hand side sample has undergone post-transfer baking at 150 $^o$C for 30 min before PMMA removal with acetone while the right hand side sample was directly soaked in a 50 $^o$C acetone bath to remove PMMA. The former method results in a visibly greater concentration of PMMA residue. c) Optical microscope image of graphene on SiO$_2$/Si, showing uniform contrast over 0.5 × 0.5 mm$^2$ area. The inset is the step height measurement by atomic force microscopy in an open region in the film exposing the underlying SiO$_2$; d) histogram of the *FWHM$_{2D}$* from Raman mapping over 200×200 μm$^2$ area (96.3% of collected data points has *FWHM$_{2D}$* of 25-35 cm$^{-1}$);

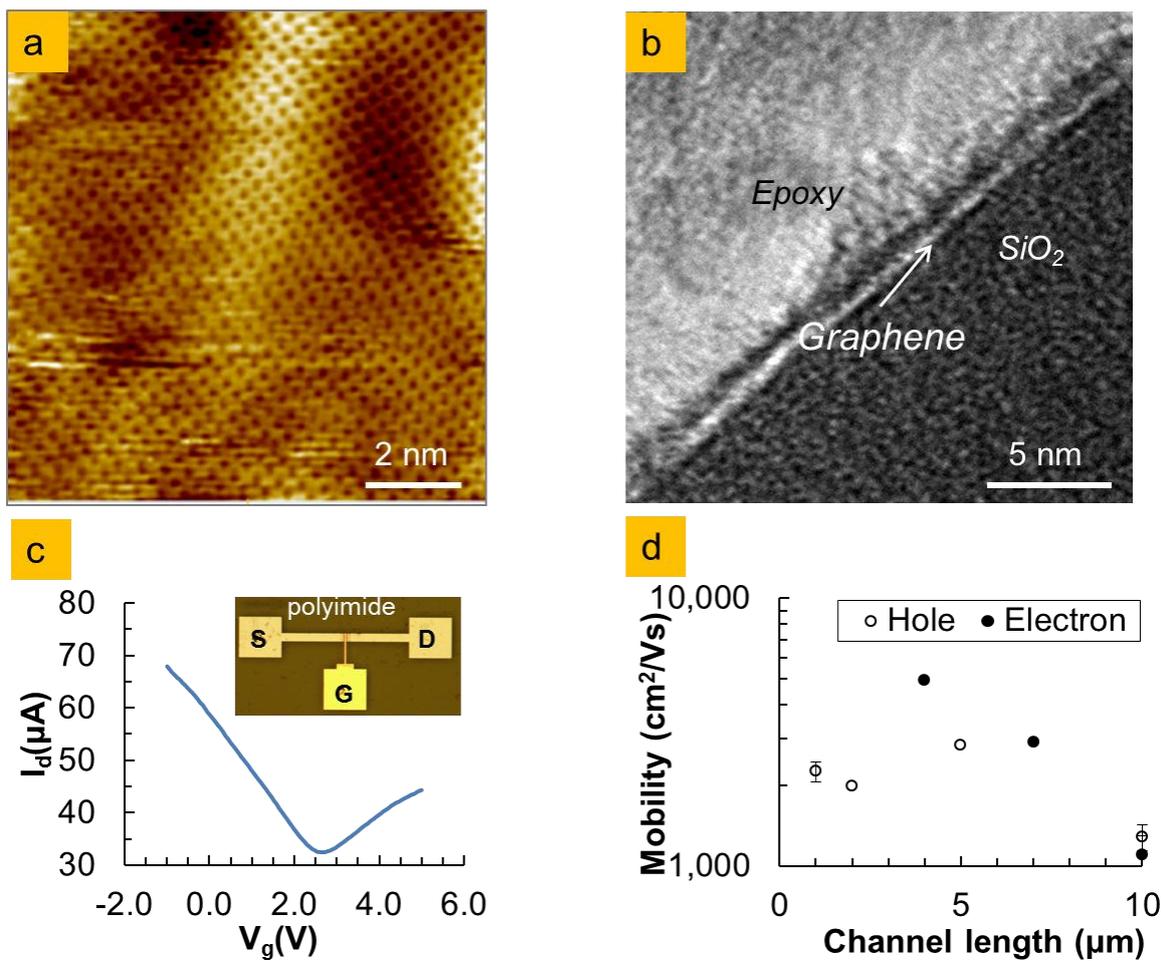

**Figure 5.** a) cross-sectional transmission electron microscope image indicates that the graphene transferred onto $SiO_2$/Si substrate is monolayer of good structural quality reflecting the b) monolayer hexagonal lattice of the as-grown graphene on copper film confirmed by scanning tunneling microscopy; c) representative $I_d$-$V_g$ curve from graphene field effect transistors with $V_d$= 100 mV and d) mobility dependence on the channel length. An average mobility is ~2,000 $cm^2$/Vs with peak value of ~4,900 $cm^2$/Vs was 5-10 times higher than most reported mobility values for graphene field effect transistors on a flexible/plastic substrate.

**Table 1.** Comparison of the reported Raman signatures of graphene obtained with various processing methods.

| Graphene obtained by various methods | | $FWHM_{2D}$ $(cm^{-1})$ | $I_{2D}/I_G$ | $I_D/I_G$ |
|---|---|---|---|---|
| This work on 100mm wafer | | 25-35 | 2-4.5 | 0-0.2 |
| CVD on deposited Cu or Ni film; (Ref [1, 4, 51-53]) | | 30-36 | 2-4 | 0.2-0.4 |
| CVD on cm-size copper foils (Ref [26, 34, 54-56]) | | 27-35 | 2-4 | < 0.2 |
| CVD/Epitaxial | on Cu (Ref [7, 9-10]) | 30-40 | 1.5-2.5 | 0.05-1 |
| | on Co,Ni (Ref [38, 42, 45-46]) | 30-40 | 0.3~3.3 | ~0.5 |
| | on Ru, Ir (Ref [39-41, 47]) | 42~46 | ~1 | 0.3-0.4 |
| | on SiC (Ref [43-44]) | 37-50 | ~1 | 0.5-1 |
| Mechanical exfoliation (Ref [17-18]) | | 25-30 | ~4 | <0.1 |

Supporting Information:

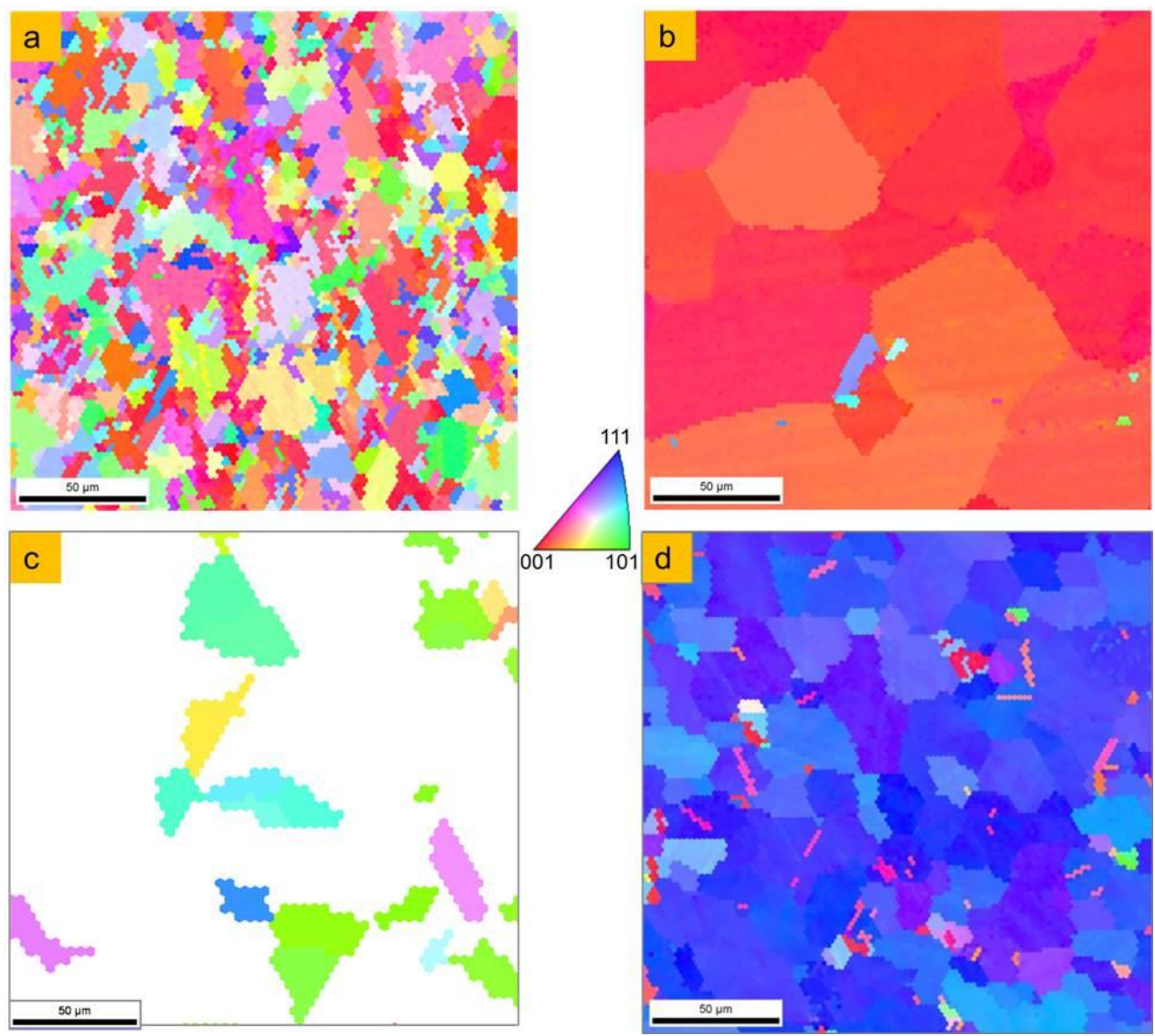

Figure S1: Electron back scattering diffraction images of a,b) copper foil and c,d) evaporated copper film before, a,c), and after the synthesis of graphene, b,d). The inverse pole figure plots were directly generated from raw data, indicating a transition from 'amorphous' (white color represents no detectable orientation) to a dominant Cu (111) film after growth.

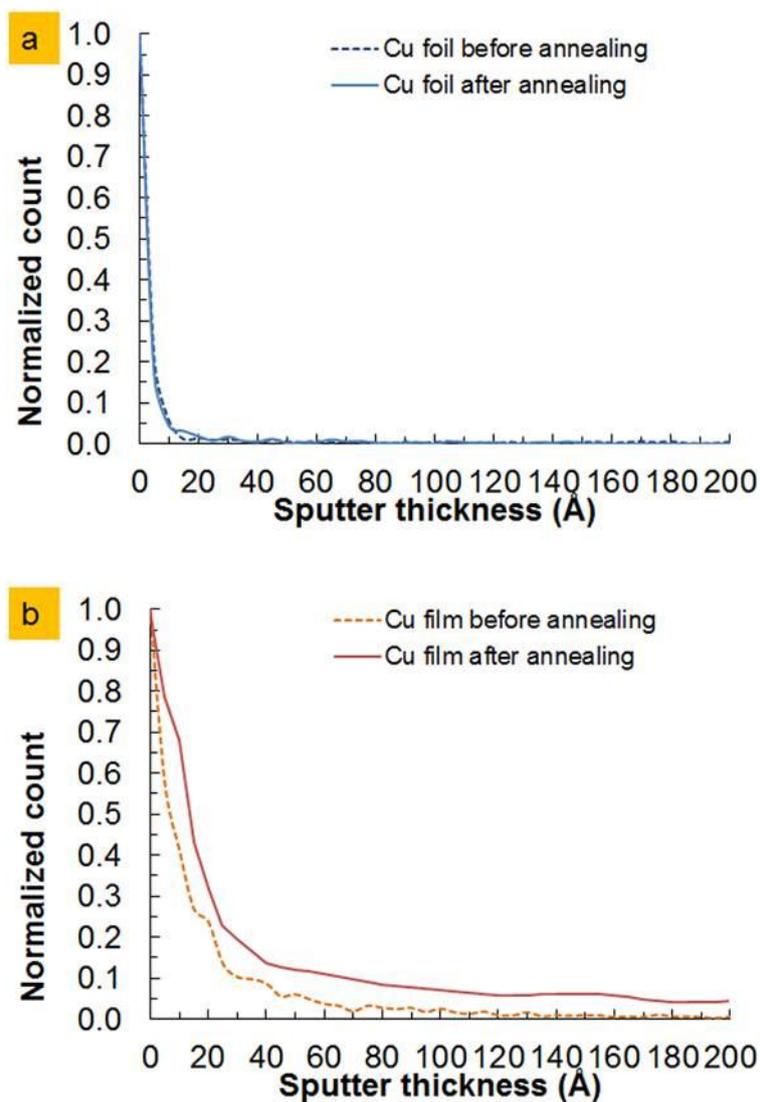

Figure S2. Time-of-flight secondary ion mass spectroscopy: depth profile of hydrogen on the surfaces of: a) copper foil, and b) evaporated copper film before and after annealing at 975 $^\circ$C, 1000 sccm $H_2$ for 5 min. There is a noticeable increase of hydrogen content on the surface of annealed copper film, whereas no difference is observed on copper foil.